\documentclass[letterpaper]{article}

\usepackage[T1]{fontenc}
\usepackage{lmodern}
\usepackage[utf8]{inputenc}

\usepackage{geometry}
\usepackage{setspace}

\usepackage[super,sort&compress]{natbib}

\usepackage{graphicx}
\usepackage{float}
\newfloat{scheme}{htbp}{los}
\floatname{scheme}{Scheme}

\usepackage{booktabs}
\usepackage{tabularx}
\newcolumntype{Y}{>{\centering\arraybackslash}X}

\usepackage{amsmath,amssymb}
\usepackage{textcomp}
\usepackage[version = 4]{mhchem}

\usepackage{authblk}

\usepackage[colorlinks=true,linkcolor=blue,citecolor=blue,urlcolor=blue]{hyperref}
\usepackage{microtype}

\author[1]{Fengxu Yang}
\author[1]{Matthew L. Schneider}
\author[2]{Weitong Chen}
\author[1]{Jack D. Evans\thanks{Corresponding author: j.evans@adelaide.edu.au}}

\affil[1]{School of Physics, Chemistry and Earth Sciences, Adelaide University, North Terrace, Adelaide, 5005, South Australia, Australia}
\affil[2]{School of Computer Science and Information Technology, Adelaide University, North Terrace, Adelaide, 5005, South Australia, Australia}

\title{Large Language Model-Assisted Systematic Study of Anomalous Adsorption Phenomena}

\date{}

\begin{document}

\maketitle

\begin{abstract}
\noindent Soft porous crystals can undergo guest-induced structural transitions producing unconventional adsorption isotherms, which are valuable for gas separation and storage applications. However, the evidence for these transitions is scattered across the literature in narrative form and has not been aggregated. Here we use a large language model pipeline to mine the literature for anomalous adsorption, extracting phase-transition events together with material, adsorbate, temperature and pressure metadata and classifying each into an established behavioural class. Benchmarked against human annotation, the extraction stage attains an $F_1$ score of $0.96$. When applied at scale, the pipeline yields a curated corpus of 547 transitions across 240 distinct materials, drawn from 200 of 4,252 screened publications, which we further link to crystallographic structures. Analysis of this database shows that the principal behavioural classes, gate-opening and breathing, are distinguished by their functional response. Gate-opening frameworks transform at a significantly lower pressure than breathing frameworks ($p = 3.8\times10^{-3}$). In contrast, no difference in pore size, void fraction, linker chemistry or metal identity was detected between the classes. Our findings argue for a functional rather than purely structural description of responsive porous materials. They also highlight the need for a community database that records the guest, temperature and pressure of every reported transition of these materials.
\end{abstract}

\section*{Keywords}

metal-organic frameworks, soft porous crystals, gate opening, breathing, adsorption isotherms, large language models, literature mining, text mining

\section{Introduction}\label{Introduction}
Porous crystalline materials, such as metal-organic frameworks (MOFs) and covalent organic frameworks (COFs), play an increasingly important role in technologies that rely on precise control of molecular adsorption, including gas storage, separations, catalysis, and drug delivery.\cite{meek_metalorganic_2011, furukawa_chemistry_2013, wu_applications_2017} Among these materials, a subset exhibits an additional and highly valuable characteristic---the ability to change their internal structure in response to external conditions. These materials, commonly referred to as soft porous crystals (SPCs), combine permanent porosity with an intrinsic structural adaptability that distinguishes them from rigid frameworks.\cite{horike_soft_2009}

The defining feature of SPCs is their sensitivity to environmental stimuli. Variations in pressure, temperature, or chemical composition can trigger abrupt structural rearrangements that directly influence adsorption behaviour.\cite{coudert_responsive_2015} Rather than filling pores gradually, these materials may remain non-porous until a critical condition is reached, after which adsorption proceeds rapidly.\cite{kitaura_porous_2003} This coupling between framework flexibility and guest uptake enables a high capacity for gas loading and release, with intrinsic management of the heat of adsorption.\cite{mason_methane_2015} This exciting potential has motivated their use in applications requiring responsive behaviours, such as sensing and controlled delivery systems.\cite{schneemann_flexible_2014} Experimentally, these adsorption-induced transformations are encoded in adsorption isotherms, where they appear as sharp steps, hysteresis loops, or other deviations from smooth uptake.\cite{bousquet_free_2012} Detecting and interpreting these features, however, is non-trivial. They frequently occur over narrow pressure and loading windows and may be masked by measurement noise or other effects, such as capillary condensation.\cite{thommes_physisorption_2015} As a result, the mechanistic significance of these anomalies is not always immediately apparent. Collecting these behaviours across many materials and conditions therefore offers a way past individual case studies, towards the identification of general design principles for soft porous crystals. Flexible frameworks are catalogued in reviews,\cite{schneemann_flexible_2014, coudert_responsive_2015} and emerging literature derived MOF databases now capture synthesis, stability and explicitly reported properties,\cite{pruyn_mof-chemunity:_2025, nandy_using_2021} but there is no structured record of which materials show anomalous adsorption. Without a rich record of these materials many basic research questions cannot be answered.

The evidence for adsorption transitions often sits in narrative text rather than in structured or consistently labelled fields. This has kept it from being analysed at scale. Large language models (LLMs) offer one promising route to exploring this information.\cite{dagdelen_structured_2024}
LLM-based extraction of structured data from the literature is now well established. Ghosh et al. recovered thermoelectric and structural properties from 10,000 full-text articles with $F_1$ scores of 0.91 and 0.82;\cite{ghosh_llm-based_2026} Li et al. reported over 95\% precision on the Organic Syntheses collection with ReactionSeek;\cite{li_reactionseek:_2026} and Pruyn et al. unified approximately 10,000 articles with 15,000 CSD structures in the MOF-ChemUnity knowledge graph.\cite{pruyn_mof-chemunity:_2025} However, these studies target explicit terms or numerical values, which reports of anomalous adsorption often lack. Evidence of a guest-induced phase transition is instead qualitative and often indirect.

Classifying the different transitions observed is even more challenging. Gate opening, breathing, multistep adsorption, and negative gas adsorption are described inconsistently across subfields and are often implied through descriptions of isotherm behaviour rather than stated explicitly.\cite{evans_four-dimensional_2020, serre_very_2002} Moreover, the terminology does not map uniquely onto the observed phenomena. A single isotherm feature, such as a sharp adsorption step, may arise from multiple underlying mechanisms, while the same mechanism may be described using different terms by different authors. Assigning a class therefore requires contextual reasoning rather than simple entity recognition.

In this work, we searched the scientific literature for reports of anomalous adsorption and assembled them into a structured, analysable database. We developed a staged LLM pipeline that extracts phase-transition events and their contextual metadata, benchmarked it against human annotation, and applied it to 4,252 publications sourced from the NIST ISODB.\cite{noauthor_nist_nodate} This approach has produced a human-verified dataset of anomalous behaviour spanning 240 distinct materials. Linking each material to its crystallographic structure and to computed pore geometry then lets us ask questions that individual studies cannot. For example, it appears the distinguishing features of these materials are functional rather than structural. The classes are separated by the pressure at which the framework transforms, a threshold that also depends strongly on the guest and the temperature, whereas no difference in static geometric or chemical descriptors was detected.

\section{Methods}\label{Method}

\subsection{Automated extraction and classification}\label{sec:extraction-methods}
We retrieved the 4,296 of the 4,644 distinct DOIs cited in NIST ISODB isotherm records (list compiled in September 2025) for which publisher-hosted, machine-readable HTML was available through institutional subscriptions.\cite{noauthor_nist_nodate} A Python workflow retrieved each page and converted it to text. Each text was then cleaned by \texttt{glm-4.5-air}\cite{glm_team_glm-45_2025} with a prompt (Supporting Information Section~S1.1) that keeps the abstract, introduction, results, discussion, conclusion and all figure captions, and removes the Experimental or Methods section, tables and their captions, standalone equations, reference lists, author and publication details and acknowledgements, without paraphrasing. The removed parts seldom describe adsorption-induced transitions, so removing them confines extraction to the sections in which transitions are usually reported and discussed. Of the 4,296 articles, 4,252 yielded usable cleaned text and were passed to extraction (Supporting Information Section~S1.1). Coverage was determined by the availability of machine-readable HTML and does not represent a complete ISODB reference list; the screened articles come mainly from four publishers (ACS 34\%, Elsevier 26\%, the Royal Society of Chemistry 21\% and Wiley 11\%).

To manage the complexity of the extraction task, we adopted a staged strategy. We first designed detailed prompts (see Supporting Information) to identify phase-transition signatures and extract the associated contextual metadata, including material identity, adsorbate species, temperature, and the relevant pressure range. Extraction used \texttt{gemini-2.5-pro}, with each cleaned article supplied whole in a single prompt (model settings in Supporting Information Section~S2).

Each material--adsorbate--temperature record was then assigned a provisional behavioural class (gate-opening, breathing, multi-step or uncertain) by ordered rules combined with LLM interpretation of its transition descriptions; the rules and the classification prompt are given in Supporting Information Section~S1.3.

Of the 1,330 records extracted from 426 publications, the 465 records from 216 publications that reported a temperature and at least one transition pressure were retained for verification; most of the remainder gave no pressure value. Every retained record was then checked against its source publication by two human reviewers working independently, and records that either reviewer flagged were resolved by cross-review. Of the 465 records, cross-review removed 32 and reclassified five, and no material name, adsorbate, temperature or pressure value required correction (Supporting Information Section~S1.4, Table~S1). All quantitative results reported below are derived exclusively from the 433 verified records.

We define the onset pressure of a record as the start pressure of its first (lowest-pressure) adsorption-induced expansion. Contractions, such as the large-pore to narrow-pore step of a breathing framework, and transitions of uncertain mode are a different event and are excluded. Therefore, a breathing record contributes the pressure at which the framework re-opens. Pressures reported relative to saturation were converted to absolute units using the saturation pressure of the guest at the measurement temperature, calculated with the \texttt{thermo} library.\cite{bell_thermo_2024} None of the 178 transitions reported as relative pressures was measured above the critical temperature of its guest, so no saturation pressure had to be extrapolated; \ce{CO2} measurements above 304~K were all reported in absolute units. Where a material has several records (different publications or temperatures), its onset pressure is the median over those records.

\begin{figure}[htbp]
    \centering
    \includegraphics[width=\linewidth]{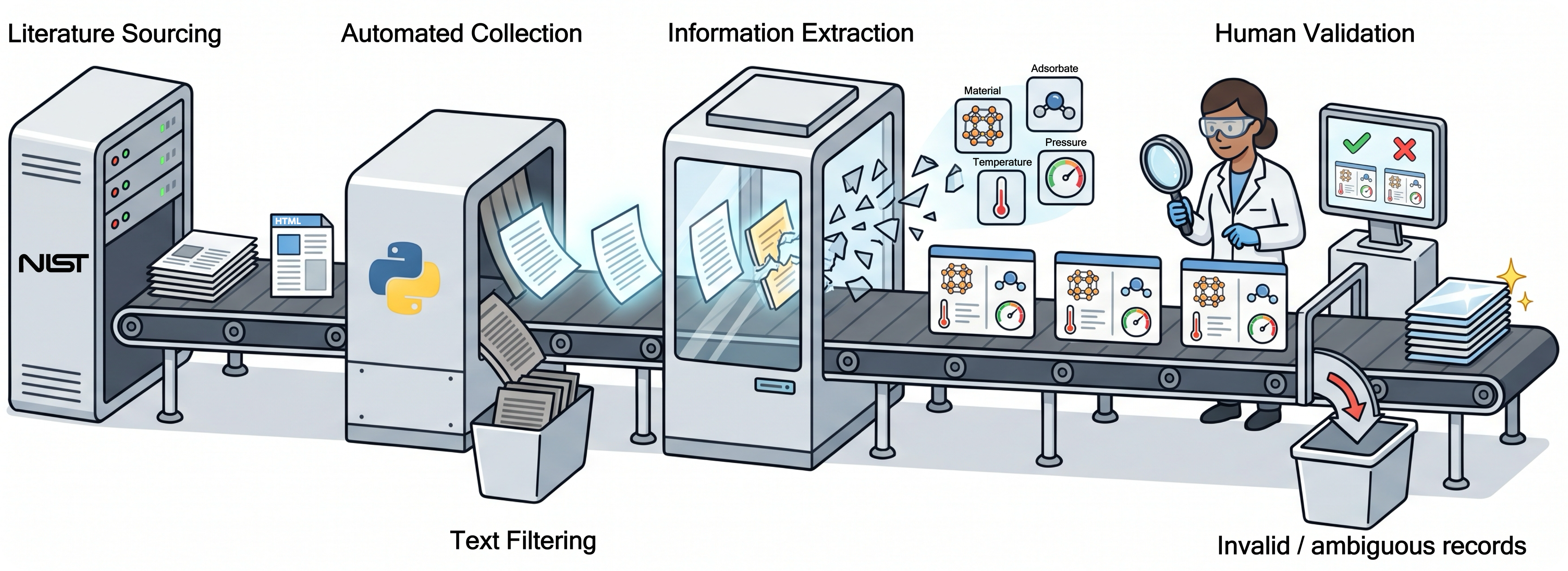}
    \caption{The staged pipeline, from a reference list to a verified corpus. Candidate publications are those cited by the NIST ISODB; full text is collected as publisher HTML, and \emph{text filtering} by a language model reduces it to the abstract, main-body sections and figure captions (Methods); the language model then extracts phase-transition signatures with their metadata and assigns a provisional behavioural class; records that report a temperature and at least one transition pressure are retained, and the authors verify every one against its source, reclassifying or discarding invalid and ambiguous records. Graphical elements in this figure were generated with Nano Banana 2 and edited by the authors; all labels and text were added and verified by the authors.}
    \label{fig:workflow}
\end{figure}

\subsection{Structure resolution and pore geometry}\label{sec:structure-methods}
To situate each extracted material within known chemical space and to enable structure--property analysis, we connected every material to a crystallographic structure and, where possible, a set of geometric pore descriptors. Material names were matched to Cambridge Structural Database (CSD)~\cite{groom_cambridge_2016} reference codes through a multi-stage procedure (Supporting Information Section~S3). In total, 185 of the 240 distinct materials were resolved to a crystallographic structure, and because the same framework is studied by multiple reports these correspond to 135 unique structures, of which 126 have pore descriptors.

Pore limiting diameters (PLD) and largest cavity diameters (LCD) were taken from the CoRE MOF 2019 database, in its November 2024 release,\cite{chung_advances_2019} when a matched structure was present in it. For matched structures absent from CoRE MOF, the same descriptors were calculated from CSD deposits using Zeo++~\cite{willems_algorithms_2012}. Supporting Information Section~S4 outlines the calculation, coverage, and validation details.

\subsection{Statistical analysis}\label{sec:statistics}
Statistical comparisons used one value per distinct material for onset pressures and one value per unique crystal structure for structural descriptors, so that frequently reported archetypes were not counted more than once. The same rule assigned a behavioural class at both levels: a distinct material, or a crystal structure taken together with every material entry linked to it, took a class only when all of its verified records carried that class, and a unit whose records carried more than one class was treated as multi-label. The gate-opening versus breathing comparisons use only units of those two classes, so multi-label, multi-step and uncertain materials and structures are excluded from them. Re-examining individual labels and links against their sources led us to exclude one breathing label that its source did not support (MOF-5) and two structure links that pointed to a different compound deposited with the same publication (Supporting Information Section~S3). Differences between the gate-opening and breathing classes were tested with the two-sided Mann--Whitney U test.\cite{mann_test_1947} Their magnitude was expressed as Cliff's $\delta$,\cite{cliff_dominance_1993} the probability that a value from one class exceeds a value from the other minus the probability of the reverse, which ranges from $-1$ to $1$. Associations between onset pressure and pore descriptors were assessed by Spearman rank correlation.\cite{spearman_proof_1904} The occurrence of categorical structural features (linker, metal node, metal and framework family) in the two classes was compared with two-sided Fisher exact tests.\cite{fisher_interpretation_1922} Intervals on Cliff's $\delta$ and on the benchmark $F_1$ scores are 95\% percentile bootstrap intervals, resampling structures within each class and publications, respectively.\cite{efron_introduction_1994} Metal and adsorbate distributions of the dataset were compared with those of the reference databases by $\chi^2$ tests,\cite{pearson_criterion_1900} with Cram\'er's $V$ reported as the effect size for the adsorbate comparison.\cite{cramer_mathematical_1946} Where a family of related tests was performed, $p$ values were adjusted with the Benjamini--Hochberg procedure to control the false discovery rate,\cite{benjamini_controlling_1995} and uncorrected values are reported alongside where relevant. Results with $p < 0.05$ are described as nominally significant. All tests were performed with SciPy 1.16.1.\cite{virtanen_scipy_2020}

\section{Results}

\subsection{Extraction and classification performance}\label{sec:extraction-performance}
We benchmarked the metadata extraction stage against a human-annotated evaluation set comprising 25 full-text publications, including five negative controls in which no anomalous adsorption behaviour was reported. None of them was used in developing the prompts, and, as for the corpus, two of the authors annotated each publication independently and reconciled their annotations by cross-review. The annotation contains 54 transition events, each a combination of material, adsorbate and temperature and so the same unit as a record of the corpus, holding 60 individual transitions with their onset pressures and pressure units. Two models were compared: the proprietary \texttt{gemini-2.5-pro} and the open-weight \texttt{Qwen3-235B-A22B}.\cite{yang_qwen3_2025} We included an open-weight model to test whether the pipeline can be run without a commercial dependency. Counting a returned transition as correct when its material, adsorbate and pressure agree with the annotation, \texttt{gemini-2.5-pro} reaches a precision of 0.95 and a recall of 0.97 ($F_1 = 0.96$, 95\% CI 0.90--1.00), and \texttt{Qwen3-235B-A22B} 0.96 and 0.88 ($F_1 = 0.92$, 0.86--0.98). The difference in $F_1$ (0.04, $-0.01$ to 0.08) is not resolved by a benchmark of this size. Both models rejected all five negative controls, so neither model assigns a transition in publications that report none. The scoring definitions and counts are given in Supporting Information Section~S2.

\begin{figure}[htbp]
    \centering
    \includegraphics[width=0.90\linewidth]{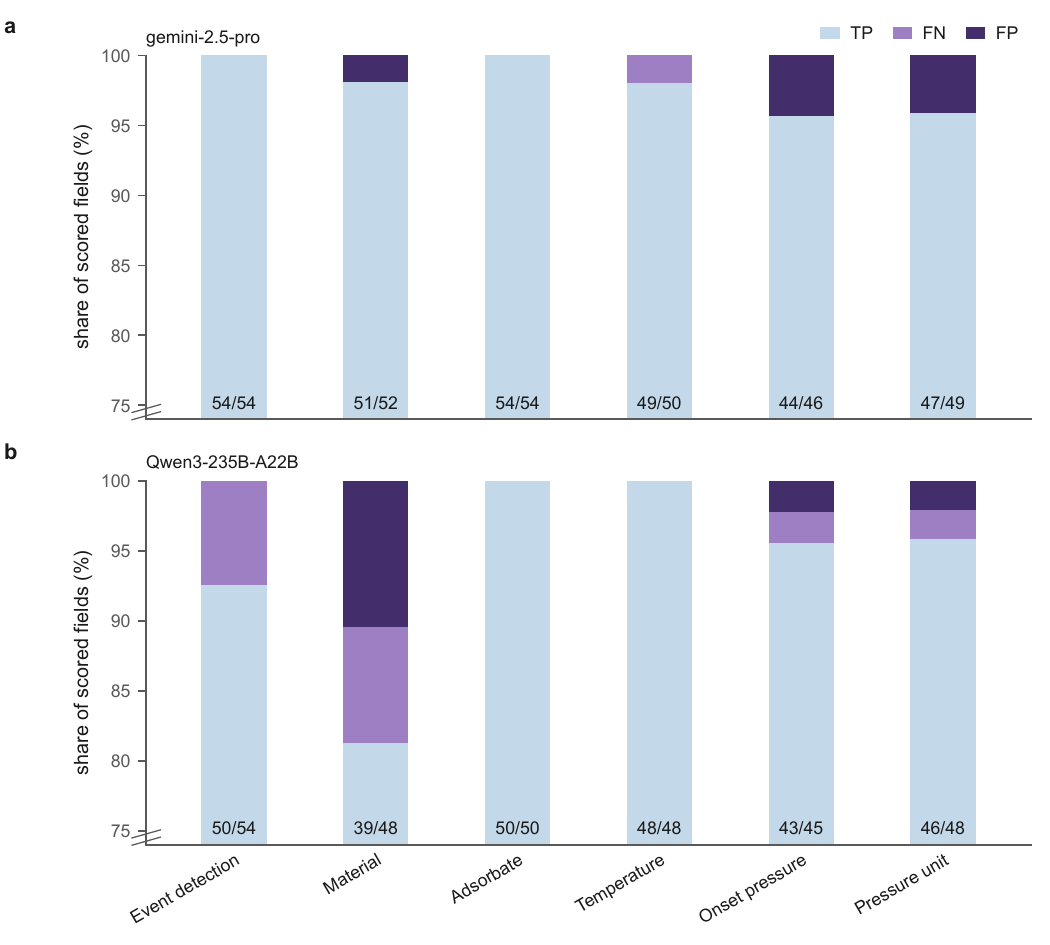}
    \caption{Field-level performance of the metadata extraction on the human-annotated evaluation set, for \textbf{(a)} \texttt{gemini-2.5-pro} and \textbf{(b)} \texttt{Qwen3-235B-A22B}. Each bar resolves one extracted field into true positives (TP, the annotated value was returned), false negatives (FN, a value is annotated and none was returned) and false positives (FP, a different value was returned); the shares sum to 100\% and the axis is truncated at 75\%, marked by the break on the baseline. Fields are compared only after their record has been matched to its annotation, so unmatched records are charged to \emph{event detection}. Material names count as agreeing when the words of one are all contained in the other. Counts and the full scoring rules are given in Supporting Information Section~S2 and Table~S2.}
    \label{fig:metrics}
\end{figure}

An aggregate score does not say which field failed, and the fields are not equally consequential (Figure~\ref{fig:metrics}). Adsorbate and temperature are recovered almost without error by both models, and the two are equivalent on the quantitative fields, recovering 44 of 46 and 43 of 45 onset pressures and 47 of 49 and 46 of 48 pressure units. They separate on completeness and on identity: \texttt{gemini-2.5-pro} detects every annotated event (54 of 54) against 50 for \texttt{Qwen3-235B-A22B}, and resolves material names in 51 of 52 records against 39 of 48. These differences are concentrated in a few publications (Supporting Information Section~S2).

We adopted \texttt{gemini-2.5-pro} for the full pipeline because of where its errors fall. Material identity is key for this study: every structure link, and therefore every comparison of structural descriptors below, uses the extracted name, and a wrong name reassigns a pore geometry to the wrong framework. Such an error can be caught only by checking the record against its source, as was done for every record of the corpus (Methods), and not by any automated stage downstream. Pressures, on which the models perform alike, are normalised to absolute units and checked in the same verification. The margin between the models is narrow, however, and rests on few publications, so the open-weight model remains a viable route at a known cost.\cite{yang_large_2026}

We also note that, while the annotated evaluation set provides a reasonably broad sample of the literature, it cannot capture the full diversity of how flexible behaviour is reported, particularly given the absence of a shared vocabulary or standardised presentation of the underlying evidence. As with any manually curated benchmark, the selection and interpretation of publications may also introduce human bias, and the resulting set is inevitably weighted towards cases in which the relevant transitions can be identified and annotated with sufficient confidence. This limitation is most apparent in class assignment: verification of the full corpus removed 32 of the 465 retained records and reclassified five (Methods), a failure mode that the evaluation set does not measure. The benchmark therefore evaluates the extraction of well-posed metadata rather than providing a definitive validation of the subsequent classification. For this reason, the full dataset underwent human verification, and an explicit \emph{uncertain} category was retained for cases in which the available evidence did not support a confident assignment.

\subsection{A curated database of anomalous adsorption behaviour}\label{sec:corpus}
Applying the pipeline to the surveyed literature yielded a curated database of 433 material--adsorbate--temperature records, containing 547 individual transitions, drawn from 200 of the 4,252 screened publications. These records describe 282 material--publication entries, which correspond to 240 distinct materials. Several archetypes are studied repeatedly. For example, MIL-53(Cr) appears in ten publications and MIL-53(Al) in nine. Each material was assigned to one of four behavioural classes on the basis of its reported isotherm signatures and the reasoning of the extraction model: \emph{gate-opening} (104 materials), \emph{breathing} (38), \emph{multi-step} (9), and an \emph{uncertain} category (61) reserved for materials whose mechanism could not be assigned unambiguously; a further 28 materials carried more than one label across independent records (Table~\ref{tab:dataset}). Negative gas adsorption was deliberately not retained as a separate class. Our classification scheme captures framework contraction and expansion, features shared by both negative gas adsorption and breathing. It does not explicitly identify the defining feature of negative gas adsorption, namely the expulsion of adsorbed gas during pressure-induced contraction.\cite{krause_pressure-amplifying_2016} Given the scarcity of reported examples, there were also too few cases to support a robust independent category. Such records were therefore assigned to the breathing or uncertain classes. Subsequent comparisons are between gate-opening and breathing, the two dominant classes; multi-label, multi-step and uncertain materials and structures are excluded from them (Methods).

\begin{table}[htbp]
\centering
\caption{Composition of the curated anomalous-adsorption dataset. Materials are counted once even when several publications report them, and are grouped by behavioural class; \emph{multi-label} denotes materials reported with more than one class across independent records. \emph{CSD structures} counts distinct Cambridge Structural Database entries, each classified by the same rule applied to every material entry linked to it, so that a structure linked to entries of more than one class is multi-label and each structure is counted once. Several materials can share one deposit (for example, different samples of MIL-53(Al)) and materials without a resolved structure contribute none, so this column does not scale with the number of materials. Three structures are not classified (Methods; Supporting Information Section~S3); the material counts are the verified labels and still include MOF-5.}
\label{tab:dataset}
\begin{tabular*}{0.68\linewidth}{@{\extracolsep{\fill}}lrr@{}}
\toprule
Behavioural class & Materials & CSD structures \\
\midrule
Gate-opening & 104 & 63 \\
Breathing    & 38  & 9  \\
Multi-step   & 9   & 4  \\
Uncertain    & 61  & 29 \\
Multi-label  & 28  & 27 \\
Not classified & -- & 3 \\
\midrule
Total        & 240 & 135 \\
\bottomrule
\end{tabular*}
\end{table}

\subsection{The anomalous dataset relative to reference MOF databases}\label{sec:distribution}
We first asked whether materials exhibiting anomalous adsorption occupy a distinctive region of chemical and structural space, by comparing the dataset against three established references: the experimentally derived structures of the QMOF database of DFT-relaxed structures (version 18)~\cite{rosen_machine_2021,rosen_high-throughput_2022}, the CoRE MOF database of experimentally reported porous frameworks~\cite{zhao_core_2025}, and the NIST ISODB adsorption-isotherm database~\cite{noauthor_nist_nodate}.

\begin{figure}[htbp]
    \centering
    \includegraphics[width=\linewidth]{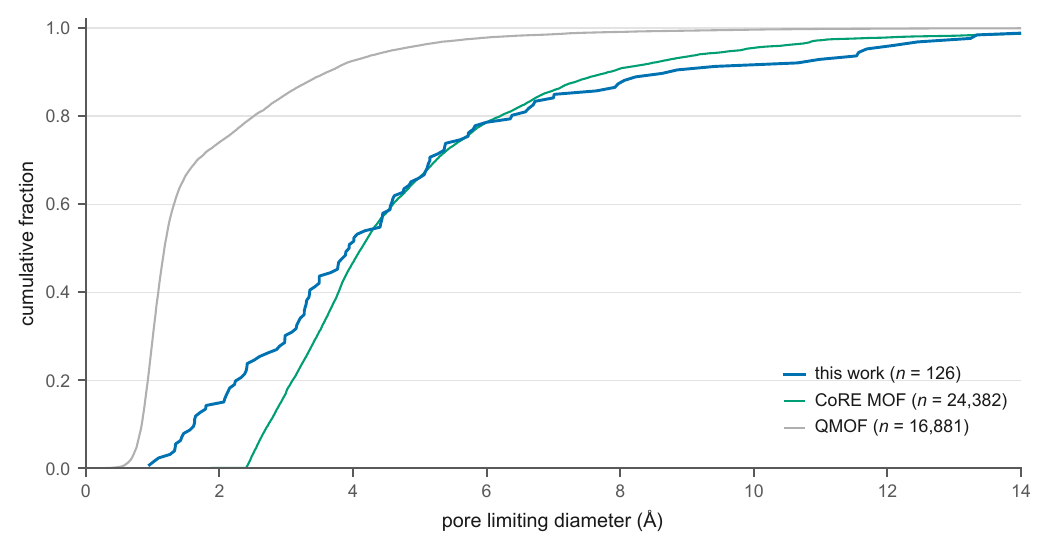}
    \caption{Pore geometry of the anomalous-adsorption dataset relative to reference databases. Cumulative distribution of the pore limiting diameter (PLD), one entry per crystal structure, shown up to 14~\AA{} (the few larger values are omitted for legibility), for this set, the CoRE MOF porous-framework database and the QMOF database of DFT-relaxed structures. The set appears open only against QMOF, which is dominated by non-porous entries; against the porous CoRE reference it is typical, its median sitting at the 45th percentile of CoRE and the 92nd of QMOF. The CoRE distribution begins at $2.4$~\AA{}, the porosity cutoff applied when that database was assembled; the structures of this set were characterised directly and are not subject to it, which is why the present curve extends below that value.}
    \label{fig:dist-geom}
\end{figure}

Whether anomalous adsorption materials can be distinguished by their pore geometry depends entirely on the choice of reference (Figure~\ref{fig:dist-geom}). Against QMOF, where the median PLD is only $1.18$~\AA{} because it is not filtered for porosity, the anomalous set (median PLD $3.94$~\AA{}) appears markedly open, at the 92nd percentile. However, when compared to CoRE MOF, which has a median PLD of $4.13$~\AA{}, the anomalous set appears typical, at the 45th percentile. Whether these frameworks appear unusually open therefore depends on the reference used. The 59 structures in our set represented in CoRE have a median PLD of $5.15$~\AA{}, compared with $2.98$~\AA{} for the 67 structures absent from CoRE. This difference could arise either from structure preparation or from the $2.4$~\AA{} porosity cutoff used in the CoRE database. To distinguish these explanations, we processed both groups using the same guest-removal and Zeo++ procedure. Among the structures calculated successfully, the medians remained $5.12$~\AA{} for the 50 represented in CoRE and $2.98$~\AA{} for the 67 absent from it. The offset therefore reflects which structures CoRE includes, rather than a difference in guest removal or PLD calculation.

\begin{figure}[htbp]
    \centering
    \includegraphics[width=0.90\linewidth]{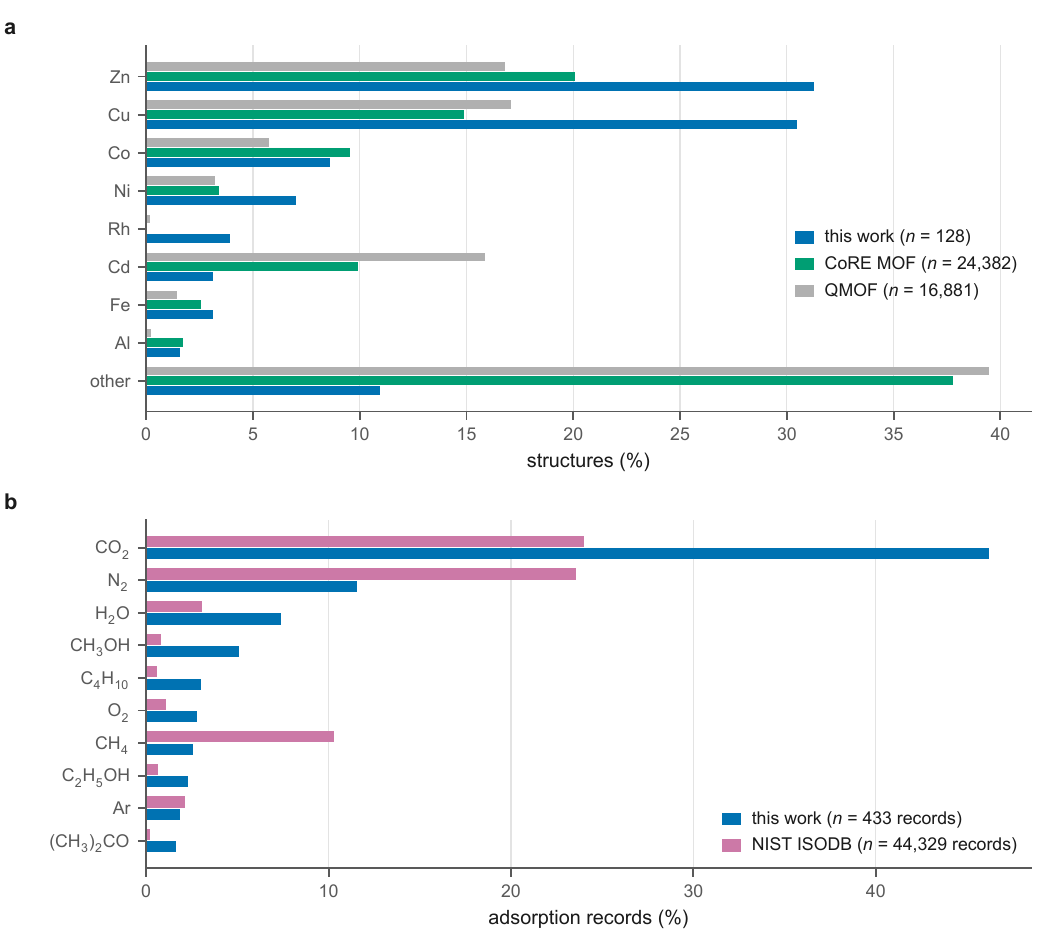}
    \caption{Chemical composition relative to background databases. \textbf{(a)} Framework-metal distribution across the anomalous set (one entry per crystal structure), QMOF, and CoRE MOF. Zn and Cu together account for 62\% of this set ($n = 128$ structures with an identified metal) against 34\% of QMOF and 35\% of CoRE ($\chi^2$ goodness of fit, $\chi^2 = 70$, $p = 4\times10^{-13}$ and $\chi^2 = 58$, $p = 4\times10^{-10}$ respectively). \textbf{(b)} Probe-adsorbate distribution in this set versus the NIST ISODB isotherm database, for the ten guests commonest in this set; \ce{CO2} and other condensable guests are over-represented and weakly interacting gases under-represented ($\chi^2 = 250$, $p = 1.6\times10^{-46}$, Cram\'er's $V = 0.07$). ISODB shares are of the 44,329 of its 54,253 isotherm records whose adsorbate could be identified, the same records used for the test. QMOF and CoRE MOF are structure-only databases and carry no adsorbate information, so ISODB is the appropriate comparator in (b).}
    \label{fig:dist-chem}
\end{figure}

In contrast to their unremarkable pore geometry, the materials are chemically concentrated (Figure~\ref{fig:dist-chem}a). Zinc and copper together account for 62\% of the linked crystal structures with an identified metal ($n = 128$), compared to only 34\% in QMOF and 35\% in CoRE. The metals outside the eight shown in Figure~\ref{fig:dist-chem}a, which together constitute roughly 40\% of both reference databases, represent just 11\% of the anomalous set. A $\chi^2$ goodness-of-fit test rejects agreement with either reference distribution ($p = 4\times10^{-13}$ versus QMOF; $p = 4\times10^{-10}$ versus CoRE). This narrow distribution reflects the discovery history of the field, in which a handful of archetypal chemistries (Zn imidazolates, Cu paddlewheel and pillared frameworks) are revisited repeatedly. As a result, the anomalous set is best understood as a chemically concentrated, rather than representative, sample of MOF space. Notably, rhodium appears in roughly 4\% of the anomalous structures (as Rh$_2$ paddlewheel analogues) yet is essentially absent from both reference databases.

The distribution of probe molecules is likewise skewed relative to the broader isotherm literature (Figure~\ref{fig:dist-chem}b). Carbon dioxide accounts for 46\% of records in our dataset but only 24\% in ISODB. Other polar or readily condensable guests (\ce{H2O}, methanol, ethanol, butane) are similarly over-represented, whereas the weakly interacting gases \ce{N2}, \ce{CH4} and especially \ce{H2} are under-represented ($\chi^2$ $p = 1.6\times10^{-46}$). This bias is physical, since condensable, strongly interacting guests are the ones most likely to induce a transition.\cite{krause_role_2021} This may also reflect research applications, as \ce{CO2} is preferentially studied on flexible frameworks for carbon capture applications.

\begin{figure}[tbp]
    \centering
    \includegraphics[width=0.90\linewidth]{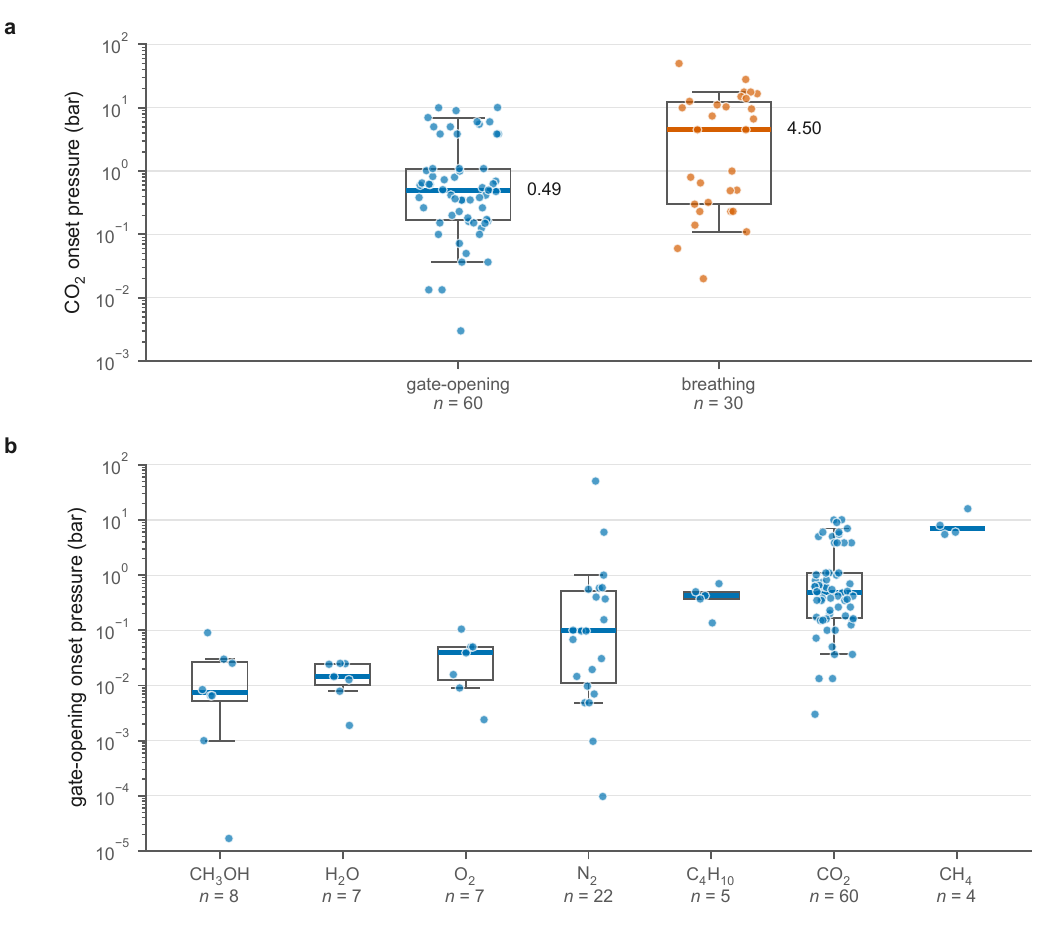}
    \caption{Transition onset pressure separates the behavioural classes. \textbf{(a)} \ce{CO2} transition onset pressure for gate-opening versus breathing frameworks, one value per material, on a logarithmic scale. Boxes give median (coloured bar, value annotated) and quartiles, whiskers the 5th--95th percentiles, and every material is plotted; gate-opening frameworks transform at markedly lower pressure (Mann--Whitney $p = 3.8\times10^{-3}$, Cliff's $\delta = -0.38$). \textbf{(b)} Onset pressure of gate-opening frameworks resolved by guest, ordered by median, on a logarithmic scale; guests reported for fewer than five materials (\ce{CH4}) are shown as points with their median only. Each guest is represented by a different set of materials; the comparison within a single host is given in the text. The threshold rises from readily condensable guests to weakly interacting ones, spanning nearly three orders of magnitude, and also shifts with temperature at fixed host and guest (see text); it is therefore a joint property of framework, guest and temperature rather than a material constant. Onset is the first adsorption-induced expansion of each record (see Methods); relative pressures were converted to absolute units against the saturation pressure of the guest at the measurement temperature.}
    \label{fig:onset}
\end{figure}

\subsection{Onset pressure is a functional discriminator}\label{sec:onset}
The clearest distinction between the classes is the pressure at which the framework transforms (Figure~\ref{fig:onset}). For carbon dioxide, gate-opening frameworks undergo their transition at a median onset pressure of $0.49$~bar (60 materials), roughly ninefold below the $4.50$~bar median of breathing frameworks (30 materials), whose onset is the pressure at which the framework re-opens. The difference is supported by a Mann--Whitney U test\cite{mann_test_1947} ($p = 3.8\times10^{-3}$) and Cliff's delta\cite{cliff_dominance_1993} ($\delta = -0.38$). Because onset pressure rises with temperature and the two classes were measured at different temperatures (median $255$~K for gate opening and $298$~K for breathing, $p = 6.6\times10^{-4}$), temperature could partly account for this difference. We therefore repeated it using only measurements between 250 and 310~K. The difference remained within this restricted range ($0.82$ versus $10.7$~bar for 30 and 18 materials, $p = 4.5\times10^{-5}$), showing that it is not solely caused by the inclusion of low-temperature measurements. Low-temperature measurements in fact narrow the gap: the cluster of breathing materials that open below 1~bar in Figure~\ref{fig:onset}a (13 of 30) consists, with one exception, of measurements near 195~K, where 1~bar approaches the saturation pressure of \ce{CO2}, and nine of these materials belong to one compositional series reported in a single publication. The separation is mechanistically reasonable as the wine-rack breathing of MIL-53 frameworks requires substantial guest-induced stress to convert the narrow-pore to the large-pore form, whereas soft, pillared gate frameworks open at a low threshold loading.\cite{neimark_stress-based_2010, kondo_gas_2012}

Onset pressure also clusters by framework family as the \ce{CO2} onset pressures of MIL-53-family frameworks, the archetypal breathing materials,\cite{loiseau_rationale_2004} lie near $6.6$~bar (median of 16 materials). By contrast, the soft porous coordination polymers of the SNU (Seoul National University),\cite{park_stepwise_2010} CID (coordination polymer with an interdigitated structure)\cite{hijikata_relationship_2011} and ELM (elastic layer-structured MOF)\cite{kondo_gas_2012} series open at a median of $0.26$~bar (7 materials, six of them below $0.65$~bar). Part of this difference may reflect measurement temperature (median 293 versus 252~K). The threshold is also strongly modulated by the adsorbate. Across gate-opening materials the median onset pressure rises from readily condensable guests (methanol and water, near $0.01$~bar) to weakly interacting ones (\ce{CH4}, near $7$~bar; Figure~\ref{fig:onset}b), although each guest is represented there by a different set of materials. The effect also holds within a single host: in the 20 gate-opening frameworks measured with more than one guest, the onset pressure spans a median factor of $9.1$ between guests. For example, in frameworks for which \ce{CO2} and \ce{CH4} were both measured, \ce{CO2} opened the gate at 9- and 17-fold lower pressure, although with $n = 2$ this pairing is qualitative.

When both the framework and the guest are fixed and only the measurement temperature varies, the onset pressure spans a median factor of $4.7$ across the 34 material--adsorbate pairs reported at more than one temperature, over a median window of only $30$~K. The direction appears uniform as the threshold rises with temperature in 31 of the 34 pairs, at roughly $0.022$ decades of pressure per kelvin, which is the sign expected for an exothermic Clausius--Clapeyron transition. At higher temperature, a higher pressure is needed to reach the chemical potential that triggers the transition, and a corresponding-states analysis of one gate-opening framework has shown the same joint dependence of the threshold on guest and temperature.\cite{sircar_corresponding_2015} Compared with the $9.1$-fold median guest effect within a single host, this temperature dependence is substantial. Transition pressure is therefore not a fixed property of a material. It depends on both the adsorbate and the measurement temperature, which should be reported with any threshold value.

Critically, this functional descriptor is largely decoupled from pore geometry. Across the dataset the transition onset pressure shows no significant correlation with the pore limiting diameter (Spearman $\rho = 0.16$, $p = 0.23$ across 62 structures and within the gate-opening and breathing classes $\rho = -0.13$, $p = 0.45$, $n = 36$, and $\rho = 0.20$, $p = 0.75$, $n = 5$). The largest cavity diameter carries a weak association ($\rho = 0.33$, $p = 0.010$), of the same sign within the computed and CoRE-sourced subsets separately ($\rho = 0.30$ and $0.10$) and so not an artefact of pooling them, but accounting for roughly a tenth of the variance in threshold pressure. The aperture a guest must cross is therefore not detectably related to the pressure at which the framework opens, while the cavity retains a modest signal. Neither approaches the separation that the classes themselves show in pressure.

\subsection{Structural descriptors do not separate the classes}\label{sec:structural}
No static structural descriptor distinguished gate-opening from breathing frameworks among the available structures. With one class per structure (Methods), the median PLD was $3.78$~\AA{} for 61 gate-opening and $4.27$~\AA{} for 8 breathing structures ($p = 0.49$, Cliff's $\delta = -0.16$, 95\% CI $-0.57$ to $0.29$), and none of the eight geometric descriptors differed after false discovery rate correction. The result does not depend on how the comparison is set up: across an alternative class rule, the exclusion of the least certain structure links and each source of pore descriptors on its own, including a uniform Zeo++ calculation for every structure, the uncorrected $p$ for PLD ranged from 0.14 to 0.73, so the uneven mix of CoRE and computed values between the classes does not create or hide a difference. Linker chemistry, metal-node geometry and metal identity were similarly uninformative: across 23 linker, node and metal features the smallest uncorrected $p$ was $0.13$. No framework-family feature reached $p < 0.05$. MIL-53, the archetypal breathing family, cannot be assessed under this rule, because every linked MIL-53 deposit also carries records of another class (Supporting Information Section~S5). With only eight breathing structures, these comparisons could detect only a large difference ($|\delta| \approx 0.6$ at 80\% power), so they show that no large structural separation was detected in this dataset, not that the classes are structurally identical. The geometric comparisons, their sensitivity analyses and the deposition state of the linked structures are given in Supporting Information Tables~S6 and S7 (Section~S4), and the chemical comparisons, unit-cell anisotropy and Figure~S1 in Section~S5.

\section{Discussion}
The database assembled here suggests a reframing of how anomalous adsorption is best organised. Gate-opening and breathing, the two classes that dominate the flexible-framework literature, are separated by their \emph{functional} response, the pressure at which the framework transforms, although their distributions overlap (Cliff's $\delta = -0.38$). In this dataset no static descriptor separated them: no difference in pore size, void fraction, cavity-to-aperture ratio, linker chemistry, metal-node geometry or metal identity was detected, although with eight breathing structures only a large difference could have been. The organising principle of these materials appears to be thermodynamic rather than compositional.

This functional view carries a practical consequence. The threshold is not a material constant, and the guest and temperature effects are each large enough to rival the difference between the classes themselves. A design target or screening criterion should therefore specify the framework, adsorbate and measurement temperature rather than assign one ``gate pressure'' to a material. Among the 10 gate-opening materials with more than one deposited structure of the same compound, the PLD differs between deposits by a median of only $0.31$~\AA{}, and by at most $1.2$~\AA{}, consistent with small variations rather than a large-scale change of framework (Supporting Information Section~S5). Only one breathing material had comparable data, too few to determine whether breathing shows a different pattern.

The composition of the dataset sharpens both the opportunity and the caveat. Chemically the set is strikingly concentrated, reflecting a discovery history built around a handful of archetypal chemistries, while geometrically the same materials are unremarkable. Much of the chemical diversity of MOF space therefore remains untested for flexibility, and generalisations from the current literature should be drawn accordingly. In general, geometric descriptors are difficult to apply here as they characterise a single structure a database happens to hold rather than the dynamics of these materials.

Several features of the corpus temper these conclusions and indicate the work that remains. Its coverage is that of the ISODB-cited articles with machine-readable HTML, so it inherits the language, publisher and topical biases of that reference list; Elsevier articles, a quarter of those screened, supplied about a tenth of the final corpus. The cleaning step removed Experimental sections and tables, and records without a reported temperature and transition pressure, about two thirds of those extracted, were not retained, so the corpus favours transitions that are quantified in the main text of an article. ISODB is both the source of the screened articles and the reference for the adsorbate distribution, so that comparison shows how reports of anomalous behaviour differ from the isotherm literature they are drawn from, not from MOF research as a whole. The skew of the probe-molecule distribution towards condensable guests is a research selection effect as much as a physical one: the apparent reluctance of weakly interacting gases to induce transitions reflects in part that they are seldom tested rather than that they invariably fail. Labels remain imperfect: a label can follow the source's own reading of an isotherm without structural evidence, as for the MOF-5 record excluded here. Structure links made through a publication DOI cannot always establish which of several deposits a material denotes; two such links were found to be wrong, and the structural comparisons are unchanged when the remaining ambiguous links are excluded. Finally, only nine breathing structures carry a single class, none of them a MIL-53 deposit, so the structural statistics of that class remain provisional.

\section{Summary and Outlook}\label{Conclusion}
We have assembled a human-verified corpus of anomalous adsorption behaviour covering 240 materials from 4,252 screened publications, and linked each material, where possible, to its crystallographic structure and computed pore geometry. We then tested geometric, chemical and functional descriptors for their ability to distinguish the principal modes of flexible response. The pressure at which a framework transforms was the most informative, and that pressure depends strongly on the guest and the temperature. Gate-opening frameworks transform at a median \ce{CO2} onset pressure of $0.49$~bar against $4.50$~bar for breathing frameworks, and the threshold of a single host varies by a median factor of $9.1$ between guests and $4.7$ across measurement temperatures.

The database is dominated by zinc and copper archetypes, so much of the flexibility latent in MOF chemical space remains unexplored. Further curation of a verified set of breathing structures is sorely needed to establish the structural features of this class. The threshold pressure depends jointly on framework, guest and temperature. Models that combine adsorbate properties, such as polarisability and boiling point, with the measurement temperature may therefore be able to predict it. Finally, the functional descriptors identified here provide concrete design targets for responsive materials in low energy separation, gas storage and sensing.

\section*{Data and Software Availability}

The curated corpus, the resolved structure links, the computed pore geometry descriptors, the extraction and classification prompts, and the analysis code that reproduces every figure are available on Zenodo at \url{https://doi.org/10.5281/zenodo.23381050}.\cite{yang_data_2026} The analysis uses Python with open-source libraries (versions listed in the repository); Zeo++ is open source; structure export used the licensed CSD Python API, and the exported structures are distributed subject to the CCDC terms of use. The publisher full texts used for extraction are not redistributed.

\section*{Supporting Information}

The Supporting Information is provided as an ancillary file with this preprint (\texttt{anc/datamining\_si.pdf}); the prompts it reproduces are also supplied there as plain-text files. The Supporting Information contains the text-cleaning, extraction and classification prompts, the classification rules and the outcome of human verification (Section~S1, Table~S1); extraction benchmark scoring and per-field outcomes (Section~S2, Table~S2); structure-resolution tiers, the class rule for structures and the withdrawn links (Section~S3, Table~S3); structure preparation, pore analysis, agreement with CoRE MOF, and the descriptor comparisons with their sensitivity analyses (Section~S4, Tables~S4--S7); additional structural comparisons (Section~S5, Figure~S1); and a description of the data and code bundle (Section~S6).

\section*{Author Contributions}

F.Y.: methodology, code development, investigation, data curation, writing -- original draft; M.L.S.: investigation, data curation; W.C.: methodology, writing -- review and editing; J.D.E.: conceptualisation, supervision, writing -- review and editing.

\section*{Competing Interests}

The authors declare no competing financial interest.

\section*{Acknowledgements}
J.D.E. is the recipient of an Australian Research Council Discovery Early Career Award (project number DE220100163) funded by the Australian Government. We thank the Phoenix HPC service at Adelaide University for providing high-performance computing resources. This research was supported by the Australian Government's National Collaborative Research Infrastructure Strategy (NCRIS), with access to computational resources provided by Pawsey Supercomputing Research Centre through the National Computational Merit Allocation Scheme.
The authors used Claude 5 (Anthropic) models to assist in drafting portions of the manuscript text and in developing the analysis codes, and Nano Banana 2 (Google DeepMind) to generate graphical elements of Figure 1. All AI-generated text and code were reviewed, revised, and validated by the authors, who take full responsibility for the content of this publication.

\bibliographystyle{nature}
\bibliography{references}

\end{document}